\documentclass[12pt,twoside,a4paper]{article}
\usepackage{psfig}

\begin{document}

\title{Constructing States and Effective Hamiltonians in Lattice QCD}
\author{
Norbert E. Ligterink
\footnote{present address: Physics Department, University of Pittsburgh (PA), U.S.A. }\\
ECT*, Villa Tambosi,\\
Strada delle Tabarelle 286, Villazzano (Trento), 38050 Italy
}
\gdef\theauthor{N.E. Ligterink}
\gdef\thetitle{Constructing States and Effective Hamiltonians in Lattice QCD}

\maketitle

\begin{abstract}
We formulate SU(3) Hamiltonian lattice QCD in terms of the plaquette variables 
and determine the relevant subspaces of the Hilbert space for the vacuum wave 
functional and its approximations. We analyze the one- and two-plaquette problems.
\end{abstract}

In an earlier paper \cite{LWB00} we formulated the fully gauge-fixed SU($N$) lattice 
gauge Hamiltonian
and studied the one-plaquette problem for several values of $N$. This paper
is restricted to the case of SU(3) lattice gauge theory. For this case I discuss
ways to go beyond the one-plaquette approximation. 

The Hamiltonian approach to lattice QCD, although more complicated, has
advantages over the Euclidean formulation. \cite{KS75,Cre83} In general, one can say that 
the Hamiltonian approach allows one to split the problem in parts and
tackle each part separately, thereby generating more understanding and
restricting the eventual computational effort. This should be formulated
consistently as restrictions on the full physical Hilbert space to a proper 
subspace, or likewise, either by formal arguments, or
from calculations. One important step in this process is the gauge fixing,
such that all the degrees of freedom are physical, and an excitation spectrum,
determined from an ab-initio many-body approach, such as RPA, is actually 
physical, and does not contain oscillations in the gauge parameter. The first 
and foremost part of the full QCD problem is the determination of the vacuum 
wave functional.

The  dimensionless QCD lattice Hamiltonian can be split into three parts:
\begin{equation}
H = 2 \sum_{\alpha} (E^a_\alpha)^2 - \frac{\lambda}{2} \sum_{\alpha} h_{\alpha M} 
+ \sum_{<\alpha,\beta>} E_\alpha^a E^a_\beta  \ \ ,
\end{equation}
where the sum over $\alpha$ runs over all plaquettes, and the sum over $<\alpha,\beta>$ runs
over all nearest-neighbour plaquette pairs. The dimensionless energy $\varepsilon$
is related to the energy $E$ for a given lattice spacing $a$ and oupling constant $g$:
\begin{equation}
\frac{g^2}{a} \varepsilon + \frac{3}{g^2 a}  =  E \ \,
\end{equation}
and the coupling constant $\lambda$ can be expressed in terms of the gauge-field
coupling constant $g$:
\begin{equation}
\lambda = \frac{1}{g^4} \ \ .
\end{equation}
If we ignore the nearest-neighbour interaction, the Hamiltonian is local and the 
solution is the product wave functional of one-plaquette wave functions. \cite{RW80}

It is important to realize that although a special unitary $3\times 3$ matrix $U$ is parametrized
by eight angles, \cite{BR65} only two are relevant for the one-plaquette problem. Since
the potential, or magnetic, term $h_M$ \cite{Wey46} only depends on the eigenvalues $x, y$ and $z$:
\begin{eqnarray}
h_M & =  &{\rm Tr} [U + U^\dagger ] \nonumber \\
&  = &   
{\rm Tr} \left[\begin{array}{ccc} x &  & \cr & y & \cr & & z \end{array}\right]+
{\rm Tr} \left[\begin{array}{ccc} x^{-1} &  & \cr & y^{-1} & \cr & & z^{-1} \end{array}\right] 
\nonumber\\ & = & x+y+z+\frac{1}{x} + \frac{1}{y} +\frac{1}{z}
\end{eqnarray}
of the plaquette
variable $U$ with the relation $x y z = 1$ it is sufficient to know 
$x = \exp i \phi_x $ and $y = \exp i \phi_y$.  In the case of several plaquettes,
we can simultaneously diagonalize multi-plaquette states with up to five 
different plaquettes, only for correlations of six and more plaquettes
simultaneous diagonalization might not be possible in special cases.

For the ground state at zero temperature there are no excitations in the 
transverse directions because the Hamiltonian does not couple to these states, so
the wave functional is constant for these angles.\footnote{The normalization is not a problem since
the group is compact. Correspondingly, the maximal-Abelian gauge is not fully Abelian because 
of correlations of sixth and higher order spoil the diagonalization in three spatial dimensions.} 
So for our analysis of the vacuum, we can restrict ourselves to
the variables $x$ and $y$, however, for symmetry reasons, we use also $z$, keeping in mind
that $z=(xy)^{-1}$.

So in terms of the eigenvalues $x,y,$ and $z$ of the plaquette variable $U$ the one-plaquette
Hamiltonian is:
\begin{equation}
h = 2 \Delta + 4 \Xi - \frac{2}{3} (\Upsilon^2  + 6 \Upsilon)
 - \frac{\lambda}{2}\left(x + y + z + \frac{1}{x} +
\frac{1}{y} + \frac{1}{z} \right)  \ \ ,
\label{ham}
\end{equation}
where we define
\begin{eqnarray}
\Xi & = & \frac{1}{x-z}(x^2 \partial_x - z^2\partial_z) +
 \frac{1}{x-y}(x^2 \partial_x - y^2 \partial_y) +
\frac{1}{y-z}(y^2 \partial_y - z^2 \partial_z)  \ \ , \nonumber\\
\Delta & = & (x\partial_x)^2 + (y \partial_y)^2 + (z \partial_z)^2\ \ , \nonumber\\
\Upsilon & = & x\partial_x + y \partial_y + z \partial_z \ \ .
\end{eqnarray}
The operators $\Delta$ and $\Upsilon^2$ are diagonal for each monomial $x^p y^q z^l$, and 
give the respective eigenvalues, $p^2+q^2+l^2$ and $(p+q+l)^2$.
The terms $\Xi$ and $\Upsilon$ are due to the curvature of the group manifold given
by the Jacobian $(x-y)(y-z)(z-x)$.
Only $\Xi$ mixes the different monomials, however,
perserves the order $p+q+l$. It is easy to see that the electric operator is consistent
with the SU(3) group,
since for any eigenstate $\phi(x,y,z)$ one can show that $(xyz)^r \phi(x,y,z)$ is an eigenstate
with the same eigenvalue, which then can be mapped on $\phi(x,y,z)$.
In the strong-coupling limit $\lambda = 0$, the solutions are the
eigenstates of the electric operator, which form an orthonormal basis on the group manifold.
Generally, we do not have
to perform any integration over the group, as the eigenstates of the electric operator
guarantees orthogonality of the basis states. Actually, all we need for an eigenvalue 
equation is a unique labeling of states, orthogonal or not, an invariant subspaces, on 
which the action of the Hamiltonian is bijective, such that a truncation to
a finite subspace is possible. A similarity transform can make the Hamiltonian hermitian if necessary. 
For the one-plaquette problem these
states are orthogonal and characterized by a label $(n,m)$ where $n>m>0$:
\begin{equation}
\langle x,y,z | n,m \rangle = \chi_{(n,m)} = \frac{ \left|  \begin{array}{ccc} x^n & x^m & 1 \cr
y^n & y^m & 1 \cr z^n & z^m & 1 \end{array} \right|  }{ \left|\begin{array}{ccc} x^2 & x & 1 \cr
y^2 & y & 1 \cr z^2 & z & 1 \end{array}\right| }\ \ ,
\label{chi}
\end{equation}
with eigenvalues $\frac{2}{3}(n^2 + m^2 - n m -3)$. The eigenvalues are degenerate for each
pair $(n,m)$ and $(n,n-m)$, which corresponds to the hermitian conjugate solutions. 
In principle, for the vacuum wave functional, the basis can be restricted to symmetric
combinations only: $\frac{1}{\sqrt{2}} (|n,m \rangle + |n,n-m\rangle )$.
There are several equivalent representations of these states.
For example, they can be expressed as polynomials \cite{McD79} of trace variables: $t_1 = x + y + z$ and 
$t_2 = x^2 + y^2 + z^2$, or in terms of loop variables: 
$l_{i,j} = (x^i + y^i + z^i)(x^j + y^j + z^j)$. In the mappings between these 
representations one makes extensive use of the relation $x y z = 1$. In order to
derive results efficiently we need to go to and fro between representations. We choose 
the representation in terms of symmetric polynomials of the eigenvalues
because of its clarity and versatility. I will return to this representation to
check signs and factors in the formulae.

\begin{table}[tb]
\begin{center}
\begin{tabular}{|c|r|c|}
\hline
$(n,m)$ &  $  \chi_{(n,m)} = \sum c_{ijk} m_{ijk},\  i+j+k=n+m-3,\  
i\geq j\geq k$  & $ \lambda_{(n,m)} $ \\
\hline
(2,1) & $  1 $  &  0 \\
(3,1) & $  m_1$  &  $ {8}/{3} $ \\
(3,2) & $  m_{1,1}$ &  $ {8}/{3} $ \\
(4,1) & $m_2 + m_{1,1}$&  $ {20}/{3} $ \\
(4,2) & $ m_{2,1} + 2 m_{1,1,1} $ & 6   \\
(4,3) & $  m_{2,2} +   m_{2,1,1}$ & $ {20}/{3}$ \\
(5,1) & $    m_{3} + m_{2,1} + m_{1,1,1} $  & $ 12$ \\
(5,2) & $   m_{3,1} + m_{2,2} + 2 m_{2,1,1}$ & ${32}/{3} $ \\
(5,3) & $   m_{3,2} + m_{3,1,1} + 2 m_{2,2,1}$ & ${32}/{3} $ \\
(5,4) & $   m_{3,3} + m_{3,2,1} + m_{2,2,2}$  & $ 12$ \\
(6,1) & $  m_{4} + m_{3,1} + m_{2,2} + m_{2,1,1}$ & $ {56}/{3} $ \\
(6,2) & $   m_{4,1} + m_{3,2} + 2 m_{3,1,1} + 2 m_{2,2,1}$& $ {50}/{3} $  \\
(6,3) & $   m_{4,2} + m_{3,3} + m_{4,1,1} + 2 m_{3,2,1} + 3 m_{2,2,2}$  & $ 16 $ \\
(6,4) & $    m_{4,3} + m_{4,2,1} + 2 m_{3,3,1} + 2 m_{3,2,2} $ & $ {50}/{3} $ \\
(6,5) & $    m_{4,4} + m_{4,3,1} + m_{4,2,2} + m_{3,3,2} $& $ {56}/{3} $ \\
\hline
\end{tabular}
\end{center}
\caption{ The eigenstates $(n,m)$ expressed in monomials, for example:
 $m_{3,2,2} = x^3 y^2 z^2 + y^3 x^2 z^2 + z^3 x^2 y^2$ and
$m_{2,2,2} = x^2 y^2 z^2$. These expressions can be reduced, using the
relation $x y z = 1$, however, for most purposes it useful to keep
the original form. The last column contains the corresponding eigenvalues.
}
\label{tabs}
\end{table}

In Table \ref{tabs} we express the eigenstates up to $n$ is six in terms of monomials.
The, generally integer, coefficients of the mapping between representations can be deduced
from group theoretical methods using Young tableaux,\cite{Ham62} but in this paper I will not refer 
to any group-theoretical result.
A general formula for an electric eigenstate is given by:
\begin{eqnarray}
\langle x,y,z | n,m \rangle  & = & \sum_{i=0}^{m-1} \sum_{j=0}^{n-1} \sum_{k=0}^{|i-j|-1}
\sigma_{j-i} z^{n+m-2-i-j} x^{\frac{i+j +|i-j|}{2} - 1 - k} y^{\frac{i+j-|i-j|}{2}+k} \nonumber \\
& = &\sum_{\begin{array}{c}i\geq j\geq k \\ i+j+k = n+m-3 \end{array}} c_{i,j,k} m_{i,j,k} \ \ ,
\end{eqnarray}
which follows from the evaluation of the determinants of Eq.~\ref{chi}. 
We used the result, which will be useful later for the evaluation of 
the $\Xi$ operator,
\begin{equation}
\frac{x^i y^j - x^j y^i}{x-y} = {\sigma_{i-j}}\ (x y)^{\min\{i,j\}} 
\sum_{k=0}^{|i-j|-1} x^{|i-j|-1-k} y^k \ \ .
\end{equation}
In all this, $\sigma$ is the sign function, with $\sigma_0=0$.

\begin{figure}[tb]
\centerline{\psfig{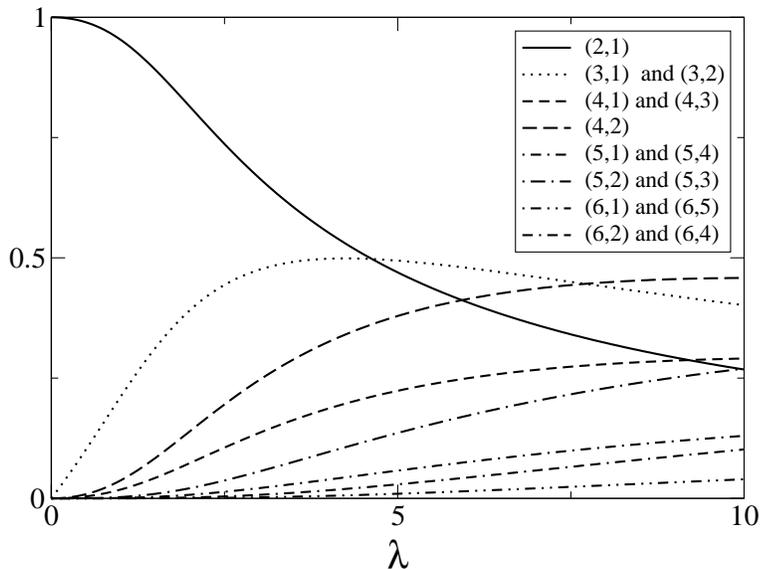}}
\caption{The occupation numbers of the lowest basis states for the one-plaquette 
ground state. Due to the symmetry between covariant and contravariant representations,
the occupation numbers for the states $(n,m)$ and $(n,n-m)$ are identical.}
\label{fig1}
\end{figure}

\begin{figure}[tb]
\centerline{\psfig{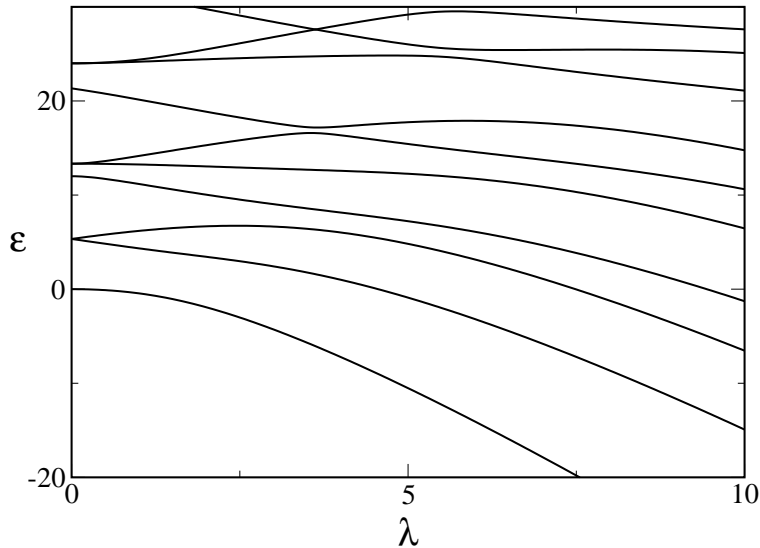}}
\caption{The low-lying energy spectrum for the one-plaquette 
ground state.}
\label{fig2}
\end{figure}

If we use this basis up to a certain number\footnote{
The Hamiltonian restricted to this subspace is hermitian.} $n$ we can solve the one-plaquette
problem for arbitrary $\lambda = g^{-4}$. Away from the strong coupling limit
a large basis is required for convergence, but with 100-200 states a good convergence is obtained
for $ \lambda < 50$. In Figure~\ref{fig1} we show the occupation number for the lowest states
as a function of the coupling constant. Clearly, a typical expansion around the strong coupling limit
with, say, 5-10 states will break down quickly, as many electric eigenstates contribute to the ground 
state. In Figure~\ref{fig2} we show the lower end of the spectrum.

Therefore, in order to go beyond the one-plaquette approximation, it is not only necessary
to construct multiple plaquette states, but also restrict the physical Hilbert space to make
any calculation feasible. A straight two-plaquette problem requires already thousands of states
for a reasonable convergence, and there will be no easy way to extend that to larger
lattices. We will return to this problem later. 

From the one-plaquette problem, we are guided to write the Hamiltonian:
\begin{equation}
H = \sum_{\alpha} h_\alpha + \sum_{<\alpha,\beta>} E^a_\alpha E^a_\beta\ \ ,
\end{equation}
where $h_\alpha$ is the Hamiltonian for the plaquette $\alpha$, which we
solved earlier. 
Although the ground-state and the low-lying states  of the one-plaquette 
problem contain many contributing basis states, it would not be too hard
to construct an effective Hamiltonian with only the eigenstates of the lower 
end of the one-plaquette spectrum. The real question is how to construct the
spatial excitations; the important multiple-plaquette correlations, which
then in its turn are used to construct the short-range interaction in the
effective Hamiltonian. In a way, part of the physics ends up in the 
coupling-constant and lattice-spacing dependence of the parameters 
of this effective Hamiltonian.

Since the operator $E^a_\alpha E^a_\beta = \Omega_{\alpha\beta} + {\cal P}_{\alpha\beta}$, 
contains a diagonal part $\Omega_{\alpha\beta}$ and a part ${\cal P}_{\alpha\beta}$ between 
plaquette $\alpha $ and plaquette $\beta $,  which brings us out of the one-plaquette 
product wave functional. We can use it as generator for spatial
excitations. For a state $|(n_\alpha,m_\alpha)\rangle |(n_\beta, m_\beta)\rangle$  the 
multiple application of the operator ${\cal P}_{\alpha\beta}$ can generate 
up to $\min\{n_\alpha,n_\beta\}-2$ 
spatial excitations. In order to see this come about we look at
a basis state in terms of its trace variables. One term may be 
written as:
\begin{equation}
w_{j_1j_2\cdots j_l} = 
\sum_{i_1,i_2,\cdots,i_{l}} U_{i_1i_{j_1}} U_{i_2i_{j_2}} \cdots  U_{i_li_{j_l}}
\label{trace}
\end{equation}
where $\{j_1j_2j_3\cdots j_l\}$ is an permutation of $\{123\cdots l\}$. In the
case of the identity: $\{j_1 j_2 j_3 \cdots j_l\} = \{1 2 3\cdots l\}$ it is a product of single
traces: $w_j = {\rm Tr}[U]^l$  and for $\{j_1 j_2 j_3\cdots j_l\} =\{l 1 2 \cdots (l-1)\}$
it yields $w_j = {\rm Tr}[U^l]$. So in the end it is not the permutation 
$\{j_1 j_2 j_3\cdots j_l\}$ that gives a unique representation, but the cocycles
in the permutation; the number of traces and each exponent. This can be labeled by 
exponents as
the partition of $l$. For the special group SU($N$) the representation in
terms of trace variables is not unique. Because $xyz=1$ every term can be expressed
in terms of two traces ${\rm Tr}[U^i] {\rm Tr}[U^j]$ since a product of three traces contains
terms like $x^i y^j z^k$ from which a common factor $(xyz)^{\min\{i,j,k\}}$ can be 
removed. This has been the greatest headache for people\cite{Leo98} studying SU(3) Hamiltonian lattice
gauge theory\footnote{For SU(2) this problem is much easier.}. It mixes also terms
of different order. As soon as one includes
more complicated states, in terms of trace variables, the abundance of redundant states
is much larger that the number of unique states. There does not seem to be an easy
way to determine the unique and complete set in terms of trace variables.

However, we will work from the other way around; we have the electric eigenstates
of the SU(3) group on one plaquette, without redundancies. We construct the corresponding 
trace variables from that, and we can now study the effect of the correlation 
operator ${\cal P}_{\alpha \beta}$. We start best from the  SU(3) identity:
\begin{equation}
\frac{1}{4} \sum_{a=1}^8 \lambda^a_{ij} \lambda^a_{kl} = \frac{1}{2} \delta_{il}\delta_{kj}
-\frac{1}{6} \delta_{ij}\delta_{kl} \ \ .
\end{equation} 
The application of the electric operator corresponds to the insertion of these
$\lambda$-matrices at the appropriate places. So acting on the right, for example, 
on the trace above Eq.~\ref{trace} at places, $r$ and $s$, we find that
\begin{equation}
\{j_1\cdots j_l\} \to \frac{1}{2} \{j_1\cdots j_{r-1} j_{s} j_{r+1} \cdots 
j_{s-1} j_{r} j_{s+1} \cdots j_l\} - \frac{1}{6} \{j_1\cdots j_l\} \ \ .
\label{perm}
\end{equation}
As simple as that.

We have the electric operator on each plaquette, which is a combination of the
permutation operator and the identity operator for each basis state, and we
have the inter-plaquette operator ${\cal P}_{\alpha \beta}$ that permutes lines between the two
plaquettes. So in the two-plaquette problem we distinguish six operators:
${\rm P}_\alpha $ and ${\rm P}_\beta$ the permutation operators of lines of plaquette $\alpha$  
and plaquette
$\beta $ respectively, $\Omega_\alpha $ and $\Omega_\beta$ the diagonal part of the electric
operator on each plaquette, ${\cal P}_{\alpha \beta}$ the operator that permutes lines of plaquette
$\alpha$ and plaquette $\beta$ , and $\Omega_{\alpha \beta}$ the diagonal part of the 
inter-plaquette operator.

Now we can make a number of important observations.
Firstly, every eigenstate on a plaquette of the electric operator can be 
expressed in terms a finite number of {\it normal-ordered} permutation
operators ${\rm P}$ without common indices on a given number of lines $l$: 
\begin{equation}
\langle U | n,m\rangle =  \sum_{l=0}^{n+m-3} \sum_{k = 0}^{[\frac{l}{2}]}
 c_{kl} :{\rm P}^k: w_{1\cdots l} \ \ ,
\end{equation}
where the normal-ordered permutation operator $:{\rm P}^k:$ can be expressed in terms of
permutations $P(i,j)$ of elements $i$ and $j$, and the inter-plaquette operator ${\cal P}$ accordingly:
\begin{eqnarray}
:{\rm P}^k:\ &  = & \sum_{i_1\neq i_2\neq \cdots\neq i_{2k}} P(i_1,i_2) P(i_3,i_4) \cdots 
P(i_{2k-1},i_{2k}) \ \ , \\
:{\cal P}_{\alpha\beta}^k: & = & \sum_{i_1\neq i_2\cdots\neq i_{k}, j_1\neq j_2\cdots\neq j_{k}}
P(i_1,j_1) P(i_2,j_2) \cdots P(i_{k},j_{k}) \ \ ,
\end{eqnarray}
where the indices $i_1,\cdots, i_k$ belong to the plaquette variables on the plaquette $\alpha $, and
$j_1,\cdots, j_k$ to the plaquette $\beta$.
Terms like $P(12)P(23) w_{123} = {\rm Tr}[U^3]$ can be reduced to 
$ w_0 + \frac{3}{2} P(12)w_{123} - \frac{1}{2} w_{123} $ $=  w_0 + \frac{1}{2} {\rm P} w_{123} -
\frac{1}{2} w_{123} $, where $w_0 = {\rm Tr}[1]$.

Secondly, the inter-plaquette operator can be normal ordered, without common indices,
and truncates at the minimal number of lines on one of the plaquettes. The normal
ordering generates additional inter-plaquette permutations:
\begin{eqnarray}
{\cal P}_{\alpha \beta} {\cal P}_{\alpha \beta} w^{(\alpha )}_{1\cdots p} 
w^{(\beta )}_{1\cdots q} & = & 
:{\cal P}_{\alpha \beta}^2 :  w^{(\alpha )}_{1\cdots p} w^{(\beta )}_{1\cdots q}  \\
 & +  & \frac{1}{p} {\cal P}_{\alpha\beta} {\rm P}_\alpha  w^{(\alpha )}_{1\cdots p} w^{(\beta )}_{1\cdots q}  
+ \frac{1}{q} {\cal P}_{\alpha\beta} {\rm P}_\beta  w^{(\alpha )}_{1\cdots p} w^{(\beta )}_{1\cdots q}
\nonumber \\
 & +  & p q w^{(\alpha )}_{1\cdots p} w^{(\beta )}_{1\cdots q} \ \ . \nonumber 
\end{eqnarray}

Lastly, the magnetic term on each plaquette can be recovered from relations like:
\begin{eqnarray}
{\rm Tr} [U] :{\rm P}^k: w_{1\cdots l} & = & \frac{l-2k}{l+1} :{\rm P}^k: w_{1\cdots l(l+1)} \\
{\rm Tr} [U^2] :{\rm P}^k: w_{1\cdots l} & = & \frac{(l-2k)(l-2k+1)}{(l+2)(l+1)}
:{\rm P}^{k+1}: w_{1\cdots l(l+1)(l+2)}
\end{eqnarray}

So every two-plaquette basis state can be represented by five labels:
\begin{equation}
\langle\{U_\alpha, U_\beta \}|k,p,l,q,r\rangle =
:{\cal P}_{\alpha \beta}^r: \ :{\rm P}_\alpha^p: \ :{\rm P}_\beta^q:\ w^{(\alpha)}_{1\cdots k} 
w^{(\beta)}_{1\cdots l} \ \ ,
\end{equation}
where $p,q,$ and $r$ are restricted by $k$ and $l$, such that $p \leq k/2$, $q \leq l/2$, and
$r \leq \min \{k,l\}$.  
Linear combinations of $:{\rm P}^p:\ w_{1\cdots k}$ span the same space as linear
combinations of $(n,m)$ with $n+m-3=k$. An important corollary of this result is that
different one-plaquette electric eigenstates do not mix due to the inter-plaquette action, since
only one eigenstate with a given eigenvalue lies within the space spanned by
$:{\rm P}^p:\ w_{1\cdots k}$.

In terms of electric eigenstates on each plaquette there would be a redundance, since, for example, 
$:{\cal P}_{\alpha\beta}^2: \langle \{U_\alpha U_\beta\} |(3,2)_\alpha (4,1)_\beta \rangle = 0$
due to cancellations among the trace variables, however, due to mixing of states with different
$n$ there is no real redundancy, if the commutation relations are worked out.
Similarly, a straightforward truncation of the
$l\leq L$ and $k \leq K$ would violated hermiticity, and it would lead to complex eigenvalues, 
since the symmetry between the covariant $(n,m)$ and the contravariant representation $(n,n-m)$ 
is broken by this truncation. It is easy to mistake the non-permuting part of Eq.~\ref{perm}
for the diagonal part of inter-plaquette part of the electric operator, however, some
care is required here since the basis in trace variables is not orthogonal.

We will illustrate most of the concepts above in a simple example. Let's consider one-plaquette
diagrams with two lines. Starting from ${\rm Tr}[U]^2 = w_{12}$ there are two diagrams:
$w_{12}$ and ${\rm P} w_{12}$. The electric operator in this case is:
\begin{equation}
E^2 = \left( {\rm P} - \frac{2}{3} \right) + 2 \frac{8}{3} \ \ ,
\end{equation}
where the last term is the diagonal part from ${\rm Tr}[U] \sum_a {\rm Tr}[\lambda^a \lambda^a U]$,
\footnote{This equals $(N^2-1)/N$ for $N=3$ times the number of matrices $U$ in the product $w$.}
and the first part is the specific case of Eq.~\ref{perm} where 
$ {\rm P} = 2 P(1,2) = P(1,2)+P(2,1)$. The
eigenvalue equation is given by:
\begin{equation}
\left( \begin{array}{cc} \frac{14}{3} & 2 \\ 2 &  \frac{14}{3} \end{array} \right)
\left( \begin{array}{c} w_{12} \\ \frac{1}{2} {\rm P} w_{12} \end{array} \right) = \lambda
\left( \begin{array}{c} w_{12} \\ \frac{1}{2} {\rm P} w_{12} \end{array} \right)  \ \ ,
\end{equation}
which leads to the eigenstates $(3,2)$ and $(4,1)$  and the eigenvalues:
$\lambda_{(3,2)} = 8/3$ and $\lambda_{(4,1)} = 20/3$ as denoted in Table~\ref{tabs}.

If we now look at the two-plaquette problem, we see that the one-plaquette electic operator
commutes with the inter-plaquette part:
\begin{equation}
[ {\rm P}_\alpha , {\rm P}_\beta ] =  [ {\rm P}_\beta, {\cal P}_{\alpha\beta} ] =
[ {\rm P}_\alpha , {\cal P}_{\alpha\beta} ] = 0 \ \ ,
\end{equation}
therefore we can work independently in the four separate subspaces of 
$\{\phi^{(\alpha)}_{(3,2)}, \phi^{(\alpha)}_{(4,1)}\} \otimes 
\{\phi^{(\beta)}_{(3,2)}, \phi^{(\beta)}_{(4,1)}\} $.
The two-plaquette electric operator has a simple action on each of these states
\begin{equation}
E^2 =  \lambda^{(\alpha)} +  \lambda^{(\beta)} +
      \frac{1}{2} \left( {\cal P}_{\alpha\beta}  -  \frac{4}{3}\right) \ \ ,
\end{equation}
where the factor $\frac{1}{2}$  is because the permuting part of the electric operator 
on the common link is: $\frac{1}{4}({\rm P}_\alpha + {\rm P}_\beta + 2 {\cal P}_{\alpha\beta})$.
On all the other links there is a single $\frac{1}{4} {\rm P}_\beta$ or $\frac{1}{4} {\rm P}_\alpha$.
For the mixed states we get the eigenvalue equation:
\begin{equation}
\left( \begin{array}{cc} 9 & 1 \\ 1 & 9 \end{array} \right)
\left( \begin{array}{c} \phi^{(\alpha)}_{(3,2)} \phi^{(\beta)}_{(4,1)} \\ 
           \frac{1}{2} {\cal P}_{\alpha\beta} \phi^{(\alpha)}_{(3,2)} \phi^{(\beta)}_{(4,1)}
 \end{array} \right) = \lambda
\left( \begin{array}{c} \phi^{(\alpha)}_{(3,2)} \phi^{(\beta)}_{(4,1)} \\ 
          \frac{1}{2}  {\cal P}_{\alpha\beta} \phi^{(\alpha)}_{(3,2)} \phi^{(\beta)}_{(4,1)}
 \end{array} \right) \ \ ,
\end{equation}
while for the symmetric states the electric operator yields:
\begin{equation}
\left( \begin{array}{ccc} \Lambda & 1 & 0 \\ 
1 & \Lambda \pm 1  & 1  \\
  0 & 1 & \Lambda    \end{array} \right)
\left( \begin{array}{c} \phi^{(\alpha\beta)}_{\Lambda} \\
           \frac{1}{2} {\cal P}_{\alpha\beta} \phi^{(\alpha\beta)}_{\Lambda} \\
\frac{1}{4} :{\cal P}^2_{\alpha\beta}: \phi^{(\alpha\beta)}_{\Lambda}
 \end{array} \right) = \lambda
\left( \begin{array}{c} \phi^{(\alpha\beta)}_{\Lambda}  \\
          \frac{1}{2}  {\cal P}_{\alpha\beta} \phi^{(\alpha\beta)}_{\Lambda} \\
\frac{1}{4} :{\cal P}^2_{\alpha\beta}: \phi^{(\alpha\beta)}_{\Lambda} 
 \end{array} \right) \ \ , \label{lambda}
\end{equation}
where $\Lambda = 38/3$ and the positive sign for $(4,1)$ states of each plaquette, 
i.e., $ \phi^{(\alpha\beta)}_{\Lambda} = \phi^{(\alpha)}_{(4,1)} \phi^{(\beta)}_{(4,1)}$,
and $\Lambda = 14/3$  and the negative sign for $(3,2)$. The two largest of the three eigenvalues
of the two-plaquette electric eigenstates for $(3,2)_\alpha(3,2)_\beta$ are $14/3$ and 
$17/3$, which correspond with the eigenvalues of coupled $(3,1)_\alpha(3,1)_\beta$ states.
A representation in terms of symmetric polynomials in $x_\alpha,y_\alpha$ and $z_\alpha$, and 
$x_\beta,y_\beta$ and $z_\beta$
reveals indeed that these are hermitian conjugate solutions.
The lowest eigenvalue $8/3$ of the $(3,2)_\alpha(3,2)_\beta$ triplet is degenerate with the 
lowest single-plaquette excitation like $(3,2)_\alpha (2,1)_\beta$, however, this corresponds 
to a rebundant solution, as the three states of Eq.~\ref{lambda} are linearly dependent for
$\Lambda = (3,2)$, contrary to the case of $\Lambda = (4,1)$.

This procedure can be carried out to construct any eigenstate of the electric
operator without generating many redundant states. Furthermore, it does not rely on 
the diagonalization of plaquette variable $U$ in the one-plaquette problem, so
it can be applied hierarchically up to any order in a multi-plaquette Hilbert space.
However, in the future we hope to map this procedure completely onto a basis of 
symmetric polynomials which should remove the last redundancies.

Eventually, as we increase the spatial extent of the states, we can restrict ourselves
to the lower energy states of the subsystems from which these states are build. Thereby 
we restrict the computational effort, which, as mentioned before, is the main problem
of an ab-initio Hamiltonian lattice QCD calculation.
Finally, I gratefully acknowledge discussions with Niels Walet during the early stages of this
work.


\end{document}